\documentclass[aps,prd,superscriptaddress,preprint,tightenlines,nofootinbib]{revtex4}
\usepackage{graphicx} 
\usepackage{dcolumn}  
\usepackage{bm}       

\begin{document}


%
%

\title{
 QUANTUM-CORRELATED MEASUREMENTS RELATED TO THE DETERMINATION OF $\gamma/\phi_3$
}

\author{J.~LIBBY\footnote{Proceedings of an invited talk at CHARM 2010, Beijing, China, Oct 21-24 2010, given on 
behalf of the CLEO Collaboration.}}

\address{Department of Physics, Indian Institute of Technology Madras, \\
Chennai 600028, Tamil Nadu, India\\
libby@iitm.ac.in}

\date{13th December 2010}

\begin{abstract}
Measurements of $D^{0}$ meson strong-phase parameters in quantum-correlated $\psi(3770)\to D^{0}\overline{D^0}$ 
decays by the CLEO collaboration are presented. These measurements play an important role in the determination of 
the unitarity triangle angle $\gamma/\phi_3$ from $B$-meson decays. Measurements of the strong-phase parameters for 
$D^{0}\to K^{0}\pi^{+}\pi^{-}$, $D^{0}\to K^{0}K^{+}K^{-}$, $D^{0}\to K^{-}\pi^{+}\pi^{0}$, and $D^{0}\to 
K^{-}\pi^{+}\pi^{+}\pi^{-}$ decays are described along with their impact on the determination of $\gamma/\phi_3$.

\end{abstract}
\maketitle

\section{INTRODUCTION}
One of the primary goals of flavor physics is to determine the angle $\gamma/\phi_3$ of the $b-d$ CKM triangle 
\cite{CKM}. Aside from being the least well known angle of the unitarity triangle (UT), it can be determined in 
tree-level processes that have negligible contributions from beyond the standard model physics, unlike most other 
constraints on the UT \cite{FITTERS}. Therefore, any disagreement between the tree-level measurement of 
$\gamma/\phi_3$ with predictions derived from other measurements is a signature of new physics.

The most promising decay to determine $\gamma/\phi_3$ at tree level is $B^{-}\to \widetilde{D}^{0}K^{-}$ where 
$\widetilde{D}^{0}$ is a $D^{0}$ or $\overline{D^0}$ decaying to the same final state \cite{GLW}.\footnote{Here and 
throughout this paper the charge-conjugate state is implied unless otherwise stated.} The sensitivity to 
$\gamma/\phi_3$ arises from the interference between the decay $B^{-}\to D^{0}K^{-}$ and the color and 
CKM-suppressed decay $B^{-}\to \overline{D^0}K^{-}$. The most precise measurements \cite{BELLE2,BABAR3} of 
$\gamma/\phi_3$ come from decays where $\widetilde{D}^{0}\to K^{0}_{S}h^{+}h^{-}$ \cite{GIRI,BONDAR1}. Here, $h$ is 
$\pi$ or $K$. Other promising $\widetilde{D}^{0}$ final states are $K^{-}\pi^{+}$, $K^{-}\pi^{+}\pi^{0}$, and 
$K^{-}\pi^{+}\pi^{+}\pi^{-}$ \cite{ADS,AS}. All these measurements depend on parameters related to the decay of the 
$D^{0}$ meson. Knowledge of the $D$-decay parameters {\it a priori} can greatly improve the determination of 
$\gamma/\phi_3$. These proceedings summarise the measurements \cite{KSHH,LOWREY} of these parameters made by the 
CLEO collaboration and estimates their impact on the determination of $\gamma/\phi_3$.  
   
\section{MEASUREMENT OF THE STRONG-PHASE PARAMETERS OF $D^{0}\to K^{0}h^{+}h^{-}$ DECAYS}
\begin{figure}[pb]
\begin{center}
\includegraphics*[width=0.75\columnwidth]{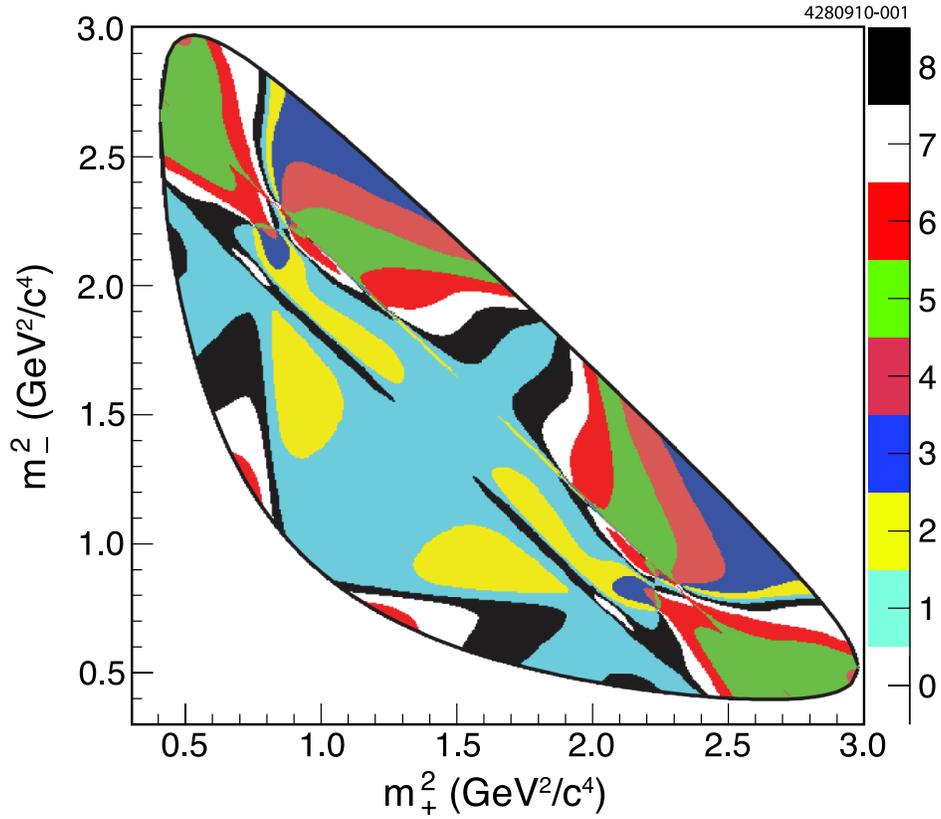}
\end{center}
\caption{Dalitz-plot binning for $D^{0}\to K^{0}_{S}\pi^{+}\pi^{-}$ in region of similar 
$\Delta\delta_D$.\label{fig:binning}}
\end{figure}

The sensitivity to $\gamma/\phi_3$ in $B^{-}\to \widetilde{D}^{0}(K^{0}_{S}h^{+}h^{-})K^{-}$ comes from studying 
differences between the $\widetilde{D}^{0}\to K^{0}_{S}h^{+}h^{-}$ Dalitz plot for both $B^{-}$ and $B^{+}$ decays. 
Current measurements of $\gamma/\phi_3$ require a model of the $\widetilde{D}^{0}\to K^{0}_{S}h^{+}h^{-}$ Dalitz 
plot, which is derived from flavor-tagged samples of $D^{0}\to K^{0}_{S}h^{+}h^{-}$. The assumptions used to 
determine the model introduce a systematic uncertainty on $\gamma/\phi_3$ which is estimated to be between 
$3^{\circ}$ and $9^{\circ}$ \cite{BELLE2,BABAR3}. This is significantly less than the current statistical 
uncertainty but it will be a limiting factor in future measurements \cite{LHCB,EPLUSEMINUS}. Therefore, it is 
desirable to perform the measurement in a model-independent manner. Such a method was proposed in Ref. \cite{GIRI} 
and has been developed significantly by Bondar and Poluektov \cite{BONDAR2}. The method requires determining yields 
in bins of the $\widetilde{D}^{0}\to K^{0}_{S}h^{+}h^{-}$ Dalitz plot for $B^{-}$ and $B^{+}$ decay, which depend 
on the $B$-decay parameters and two new parameters $c_i$ and $s_i$, which are the amplitude-weighted averages over 
the bin of the cosine and sine of the difference in strong-phase difference, $\Delta\delta_{D}$, between 
Dalitz-plot points $(m_{-}^{2},m_{+}^{2})$ and $(m_{+}^{2},m_{-}^{2})$. Here $m_{\pm}$ is the invariant-mass of the 
$K^{0}_{S}h^{\pm}$ pair. It can be shown \cite{BONDAR2,KSHH} that between 80\% to 90\% of the statistical 
sensitivity to $\gamma/\phi_3$ of the unbinned method can be obtained by choosing bins corresponding to equal 
intervals of $\Delta\delta_{D}$ according to an amplitude model. An example of such a binning is shown in 
Fig.~\ref{fig:binning}.

The values of $c_i$ and $s_i$ can be measured in quantum-correlated $D^{0}\overline{D^0}$ decays of the 
$\psi(3770)$. The $D^{0}\overline{D^0}$ are produced in a $C=-1$ state. Therefore, if one $D$ meson decays to a 
$CP$-eigenstate the other $D$-meson is in the opposite $CP$-eigenstate. The difference between $CP$-even and 
$CP$-odd tagged Dalitz plots in each bin is related to the $c_i$ parameters. In addition, the Dalitz plot of 
quantum-correlated events where both $D$-mesons decay to $K^{0}_{S}h^{+}h^{-}$ is sensitive to both $c_i$ and 
$s_i$. The strong-phase parameters for the decay $D^{0}\to K^{0}_{L}h^{+}h^{-}$ ($c^{\prime}_{i}$ and 
$s^{\prime}_{i}$) are closely related to $c_i$ and $s_i$ such that using decays of the type $K^{0}_{S}h^{+}h^{-}$ 
$vs.$ $K^{0}_{L}h^{+}h^{-}$ greatly improve the precision on $c_i$ and $s_i$.

The CLEO-c experiment \cite{CLEOC} collected $e^{+}e^{-}\to\psi(3770)\to D\bar{D}$ data corresponding to an 
integrated luminosity of 818~pb$^{-1}$. The fact that all particles arise from $D$-meson decay in the final state 
leads to both $D$ mesons being reconstructed exclusively with high efficiency and purity. 
For $D^{0}\to K^{0}\pi^{+}\pi^{-}$ ($D^{0}\to K^{0}K^{+}K^{-}$) decay the numbers of CP-tagged and 
$K^{0}h^{+}h^{-}$ $vs.$ $K^{0}h^{+}h^{-}$ candidates selected are 1661 and 1674 (219 and 335), respectively. 

A maximum-likelihood fit is performed to the bin yields of the $CP$-tagged and $K^{0}h^{+}h^{-}$ $vs.$ 
$K^{0}h^{+}h^{-}$ events to extract $c^{(\prime)}_{i}$ and $s^{(\prime)}_{i}$. 
The  results are presented in detail elsewhere \cite{KSHH}. 
The values of $c_i^{(\prime)}$ and $s_i^{(\prime)}$ are determined for several binning variations for both 
$D^{0}\to K^{0}_{S}\pi^{+}\pi^{-}$ and $D^{0}\to K^{0}_{S}K^{+}K^{-}$. These binnings allow flexibility given 
different scenarios for the amount of $B$ data and the background environment. The measured values of $c_i$ and 
$s_i$ are found to be in reasonable agreement with the values predicted by the amplitude models presented in Refs. 
\cite{BELLE2,BABAR2}. The largest systematic uncertainties arise from the modelling of the background. However, 
none of these measurements are systematically limited.

\section{MEASUREMENT OF THE COHERENCE FACTOR AND STRONG PHASE DIFFERENCES IN $D^{0}\to K^{-}\pi^{+}\pi^{0}$ and 
$D^{0}\to K^{-}\pi^{+}\pi^{+}\pi^{-}$}

The rate of decays $B^{-}\to \widetilde{D}^{0}(K^{+}\pi^{-})K^{-}$ is particularly sensitive to $\gamma/\phi_3$ 
because the two interfering amplitudes are of similar size due to the doubly-Cabibbo suppressed (DCS) $D^{0}$ decay 
coming from the favored $B^{-}$ amplitude \cite{ADS}. The rate depends not only on $\gamma/\phi_3$ but on the 
strong-phase difference between the Cabibbo-favored and DCS $\widetilde{D}^{0}\to K^{+}\pi^{-}$ decays. The 
measurement of this parameter by the CLEO collaboration is described elsewhere in these 
proceedings \cite{ASNER2}.

Via the same mechanism there is potential sensitivity to $\gamma/\phi_3$ from $B^{-}\to \widetilde{D}^{0}K^{-}$, 
where  $\widetilde{D}^{0} \to K^{+}\pi^{-}\pi^{0}$ or $\widetilde{D}^{0} \to K^{+}\pi^{-}\pi^{-}\pi^{+}$ \cite{AS}.
These modes have significantly larger branching fractions than $\widetilde{D}^{0}\to K^{+}\pi^{-}$ \cite{PDG}. 
However, the dynamics are more complicated because there is variation of the strong-phase difference over the 
multi-body phase-space. This leads to the introduction of a new parameter referred to as the coherence factor $R_F$ 
($F=K\pi\pi^{0}$ or $K3\pi$), which multiplies the interference term sensitive to $\gamma/\phi_3$. The value of 
$R_F$ can vary between zero and one. If there is only a single intermediate resonance or a few non-interfering 
resonances the coherence factor will be close to one and the decay will behave just like $\widetilde{D}^{0}\to 
K^{+}\pi^{-}$. If there are many overlapping intermediate resonances the coherence factor will tend toward zero, 
limiting the sensitivity to $\gamma/\phi_3$. However, even if there is limited sensitivity to the phases when 
$R\sim 0$ there is enhanced sensitivity to the magnitude of the amplitude ratio between the $B^{-}\to D^{0}K^{-}$ 
and $B^{-}\to \overline{D^0}K^{-}$ decays; improved knowledge of this parameter will then lead to better overall 
sensitivity to $\gamma/\phi_3$ in a global fit to all $B^{-}\to \widetilde{D}^{0}K^{-}$ decays.

\begin{figure}[pb]
\includegraphics*[width=0.75\columnwidth]{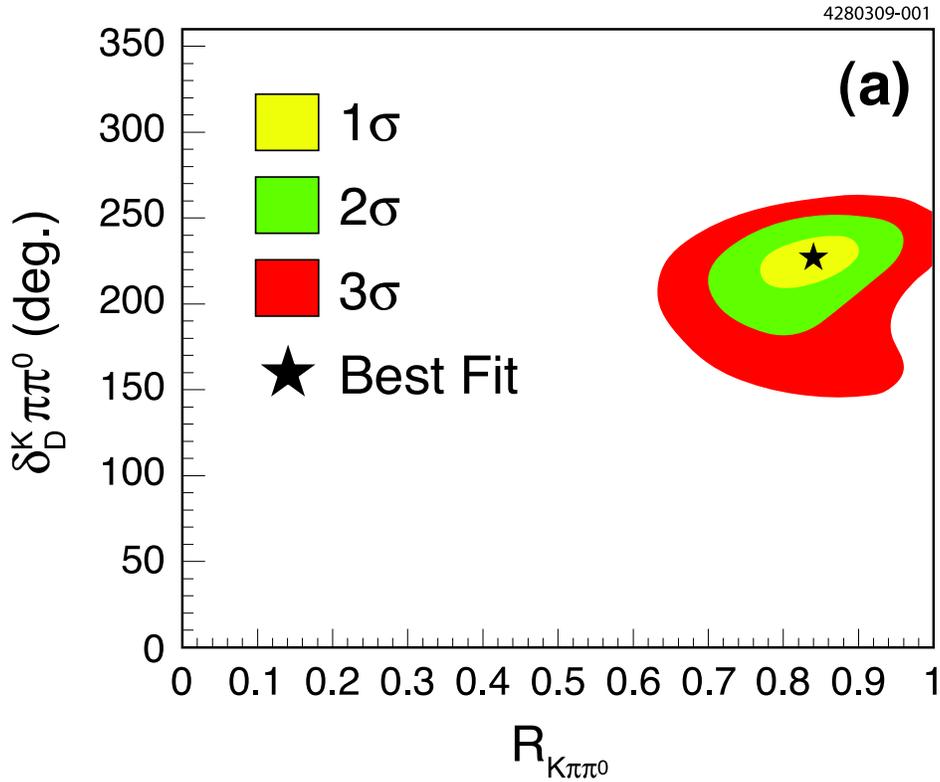}
\caption{The $1\sigma$, $2\sigma$, and $3\sigma$ allowed regions of ($R_{K\pi\pi^{0}}$,$\delta_{D}^{K\pi\pi^{0}}$) 
parameter space.\label{fig:coherence}}
\end{figure}

The values of $R_F$ and the average-strong phase difference $\delta_{D}^{F}$ have been measured by CLEO-c 
\cite{LOWREY}. Sensitivity comes from the quantum-correlated $D^{0}\overline{D^0}$ events with $F$ tagged by either 
$CP$-eigenstates or $K^{-}\pi^{+}$, $K^{-}\pi^{+}\pi^{0}$, and $K^{-}\pi^{+}\pi^{+}\pi^{-}$, where the tag kaon 
charge is the same as the signal. A $\chi^{2}$ fit to the yields gives: $R_{K\pi\pi^{0}}=0.84\pm 0.07$, 
$\delta_{D}^{K\pi\pi^{0}}=(227^{+14}_{-17})^{\circ}$, $R_{K3\pi}=0.33^{+0.26}_{-0.23}$, and 
$\delta_{D}^{K3\pi}=(114^{+26}_{-23})^{\circ}$. Figure~\ref{fig:coherence} shows the $1\sigma$, $2\sigma$, and 
$3\sigma$ regions of $(R_{K\pi\pi^{0}},\delta_{D}^{K\pi\pi^{0}})$ parameter space; the coherence of $D^{0}\to 
K^{-}\pi^{+}\pi^{0}$ is clearly observed. The impact of these results on the measurements of $\gamma/\phi_3$ is 
discussed in the following section. 

\section{IMPACT OF RESULTS ON THE MEASUREMENT OF $\gamma/\phi_3$}
\label{sec:impact}
The determination of the $c_i$ and $s_i$ in quantum-correlated $D$-decay allows the measurement of $\gamma/\phi_3$ 
without a model induced systematic uncertainty. However, this is replaced by uncertainty due to the limited 
statistics used to measure $c_i$ and $s_i$ at CLEO-c. This uncertainty is estimated to be between $1.7^{\circ}$ and 
$3.9^{\circ}$ ($3.2^{\circ}$ and $3.9^{\circ}$) depending on the binning of the $D^{0}\to K^{0}_{S}\pi^{+}\pi^{+}$ 
($D^{0}\to K^{0}_{S}K^{+}K^{-}$) Dalitz plot. The systematic uncertainty on $\gamma/\phi_3$ can be reduced by about  
a factor of three if BES-III collects $10~\mathrm{fb}^{-1}$ of integrated luminosity at the $\psi(3770)$ resonance 
\cite{BESPROC}. This would reduce the uncertainty to the order of $1^{\circ}$, which would not only be adequate for 
LHCb but also for future higher luminosity facilities \cite{EPLUSEMINUS,LHCBUP}.    

The impact of the measurements of $R_F$ and $\delta_{D}^{F}$ at LHCb is evaluated using the yield estimates for 
$B^{-}\to \widetilde{D}^{0}(K^{\pm}\pi^{\mp})K^{-}$ and $B^{-}\to 
\widetilde{D}^{0}(K^{\pm}\pi^{\mp}\pi^{+}\pi^{-})K^{-}$ decays in a dataset corresponding to $2~\mathrm{fb}^{-1}$ 
of integrated luminosity at a center-of-mass energy of $14~\mathrm{TeV}$ \cite{LHCBROAD}. In addition, the yield of 
$B^{-}\to \widetilde{D}^{0}(K^{\pm}\pi^{\mp}\pi^{0})$ is assumed to be half that of $B^{-}\to 
\widetilde{D}^{0}(K^{\pm}\pi^{\mp}\pi^{+}\pi^{-})K^{-}$ with the same level of background reflecting the 
difficulties associated with $\pi^{0}$ reconstruction in the hadronic environment. The sensitivity to 
$\gamma/\phi_3$ from LHCb data alone is $9.7^{\circ}$. Including the CLEO-c constraints on $R_F$ and 
$\delta_{D}^{F}$ this improves to $7.5^{\circ}$. The introduction of the CLEO-c constraints is equivalent to $70\%$ 
more LHCb data. This clearly illustrates the power of quantum-correlated measurements in aiding the determination 
of $\gamma/\phi_3$. BES-III data could lead to at least a further 10\% reduction of the uncertainty on 
$\gamma/\phi_3$.

In conclusion, the first quantum-correlated measurements of strong-phase parameters of $D$-decay at CLEO-c have 
been presented and their positive impact on the determination of $\gamma/\phi_3$ at LHCb has been illustrated. 
Further improvements are possible by exploiting the larger sample of quantum-correlated decays that will be 
available at BES-III. Furthermore, there are other $\widetilde{D}^{0}$ decay modes of interest to the measurement 
of $\gamma/\phi_3$ for which the strong-phase parameters have yet to be determined: 
$K^{0}_{S}\pi^{+}\pi^{-}\pi^{0}$, $\pi^{+}\pi^{-}\pi^{0}$, $K^{0}_{S}K^{\pm}\pi^{\mp}$, and 
$K^{+}K^{-}\pi^{+}\pi^{-}$.  

\section*{Acknowledgments}
I would like to thank the CHARM 2010 organisers for their financial assistance that allowed my participation in an 
excellent conference.



\begin{thebibliography}{0}    

\bibitem{CKM} N. Cabibbo, {\it Phys. Rev. Lett.} {\bf 10}, 531 (1963); 
M. Kobayashi and T. Maskawa, {\it Prog. Theor. Phys.} {\bf 49}, 652 (1973).

\bibitem{FITTERS} CKMfitter Group, (A.~H\"{o}cker {\it et al.}), {\it Eur. Phys. J. C} {\bf 21}, 225 (2001); 
CKMfitter Group (J.~Charles {\it et al.}), {\it Eur. Phys. J. C.} {\bf 41}, 1 (2005), and updates at 
http://ckmfitter.in2p3.fr/; M. Ciuchini {\it et al.}, {\it J. High Energy Phys.} 0107 (2001) 013; UTFit 
Collaboration (M.~Bona {\it et al.}), {\it J. High Energy Phys.} 081 (2006) 10, and updates at 
http://www.utfit.org/.

\bibitem{GLW} M.~Gronau and D.~Wyler, {\it Phys. Lett.  B} {\bf 265}, 172 (1991);
M.~Gronau and D.~London, {\it Phys. Lett.  B} {\bf 253}, 483 (1991).

\bibitem{BELLE2} Belle Collaboration (A.~Poluektov {\it et al.}), {\it Phys. Rev. D} {\bf 81}, 112002 (2010).

\bibitem{BABAR3} {\it BABAR} Collaboration,  (P. del Amo Sanchez  {\it et al.}), {\it Phys. Rev. Lett.} {\bf 105}, 
121801 (2010).

\bibitem{GIRI} A. Giri, Y. Grossman, A. Soffer, and J. Zupan, {\it Phys. Rev. D} {\bf 68}, 054018 (2003).

\bibitem{BONDAR1} A. Bondar, {\it Proceedings of BINP Special Analysis Meeting on Dalitz Analysis}, 24-26 Sep. 
2002, unpublished.

\bibitem{ADS} D. Atwood, I. Dunietz, and A. Soni, {\it Phys. Rev. Lett.} {\bf 78}, 3257 (1997); D. Atwood, I. 
Dunietz, and A. Soni, {\it Phys. Rev. D} {\bf 63}, 036005 (2001).

\bibitem{AS} D.~Atwood and A.~Soni, {\it Phys. Rev. D} \textbf{68}, 033003 (2003).

\bibitem{KSHH} CLEO Collaboration  (J. Libby {\it et al.}), arXiv:1010.2817 [hep-ex], accepted by Phys. Rev. D.

\bibitem{LOWREY} CLEO Collaboration (N. Lowrey {\it et al.}), {\it Phys. Rev. D} {\bf 80}, 031105 (2009).

\bibitem{LHCB} J. Libby, CERN-LHCb-2007-141; V.~Gibson, C.~Lazzeroni, and Y.-Y.~Li, CERN-LHCb-2008-028.

\bibitem{EPLUSEMINUS} T.~Aushev {\it et al.}, KEK-Report 2009-12; M. Bona {\it et al.}, SLAC-R-856.

\bibitem{BONDAR2} A. Bondar and A. Poluektov, {\it Eur. Phys. J. C} {\bf 47}, 347 (2006); A. Bondar and A. 
Poluektov, {\it Eur. Phys. J. C} {\bf 55}, 51 (2008).

\bibitem{CLEOC} CLEO Collaboration  (Y.~Kubota {\it et~al.}), {\it Nucl. Instrum. Methods Phys. Res., Sect. A} 
\textbf{320}, {66} ({1992}); D.~Peterson {\it et~al.}, {\it Nucl. Instrum. Methods Phys. Res., Sect. A} 
\textbf{478}, {142} ({2002}); M.~Artuso {\it et~al.}, {\it Nucl. Instrum. Methods Phys. Res., Sect. A} 
\textbf{554}, {147} ({2005}).

\bibitem{BABAR2} {\it BABAR} Collaboration (B. Aubert {\it et al.}), {\it Phys. Rev. D} {\bf 78}, 034023 (2008).

\bibitem{PDG} Particle Data Group (K. Nakamura {\it et al.}), {\it J. Phys. G} {\bf 37}, 075021 (2010).

\bibitem{ASNER2} D. Asner, these proceedings.

\bibitem{BESPROC} H. Liu, these proceedings.

\bibitem{LHCBUP} LHCb Collaboration, CERN-LHCb-2008-019.

\bibitem{LHCBROAD} LHCb Collaboration (B. Adeva {\it et al.}), LHCb-PUB-2009-029. 

\end{thebibliography}
\end{document}